\documentclass[aps,pra,twocolumn,showpacs,floatfix]{revtex4-1}
	
\usepackage{amsmath}
\usepackage{txfonts}
\usepackage{microtype}
\usepackage{graphicx}
\usepackage{color}
\usepackage{ulem}

\begin{document}
\title{Strong-field ionization via high-order Coulomb corrected  strong-field approximation}

\author{Michael Klaiber}\email{klaiber@mpi-hd.mpg.de}
\author{Jiri Danek}
\author{Enderalp Yakaboylu}
\author{Karen Z. Hatsagortsyan}\email{k.hatsagortsyan@mpi-hd.mpg.de}
\author{Christoph H. Keitel}
\affiliation{Max-Planck-Institut f\"ur Kernphysik, Saupfercheckweg 1, 69117 Heidelberg, Germany}

\date{\today}

\begin{abstract}
Signatures of the Coulomb corrections  in the photoelectron  momentum distribution during laser-induced ionization of atoms or ions in tunneling and multiphoton regimes are investigated analytically in the case of an one-dimensional problem. High-order Coulomb corrected strong-field approximation is applied, where the exact continuum state in the S-matrix is approximated by the eikonal Coulomb-Volkov state including the second-order corrections to the eikonal. Although, without high-order corrections our theory coincides with the known  analytical R-matrix (ARM) theory, we propose a simplified procedure for the matrix element derivation. Rather than matching the eikonal Coulomb-Volkov wave function with the bound state as in the ARM-theory to remove the Coulomb singularity,  we calculate the matrix element via the saddle-point integration method as by time as well as by coordinate, and in this way avoiding the Coulomb singularity.  The momentum shift in the photoelectron momentum distribution with respect to the ARM-theory due to high-order corrections is analyzed for tunneling and multiphoton regimes. The relation of the quantum corrections to the  tunneling delay time is discussed.
\end{abstract}

\maketitle

\section{Introduction}

In strong field ionization process of atoms and molecules the Coulomb field of the atomic core plays a significant role for electron dynamics in the continuum and for asymptotic photoelectron  momentum distribution (PMD), see e.g. \cite{Blaga_2009,Wolter_2015x}. Different schemes of attosecond spectroscopy \cite{Corkum_2007,Krausz_2009} rely on PMD to derive information on time-resolved atomic dynamics. Hence, accurate description of Coulomb effects is of paramount importance for the strong-field theory. One of the main analytical tools in the strong-field theory is the, so-called, strong field approximation (SFA) \cite{Keldysh_1965,Faisal_1973,Reiss_1980}. In the standard SFA the effect of the Coulomb field of the atomic core for the continuum electron is neglected, describing it via the Volkov wave function \cite{Volkov_1935}, corresponding to the free electron in a plane laser field. Although including the effect of the Coulomb field of the atomic core by a perturbative approach in the standard SFA as a recollision was very insightful, providing explanation for the nonsequential double ionization \cite{Becker_2002}, high-order harmonic generation \cite{Agostini_2004}, and recently also for the low-energy structures  \cite{Moller_2014,Becker_2015}, the quantitative description of fine interference structures in PMD, see e.g. \cite{Huismans_2011}, requires more accurate theory accounting Coulomb field effects nonperturbatively.

  Coulomb corrected  SFA (CCSFA) has been developed in \cite{Popruzhenko_2008a,Popruzhenko_2008b}, where the electron continuum state in the SFA amplitude is approximated by the eikonal Coulomb-Volkov state. The latter describes the electron in the laser and Coulomb fields, using eikonal approximation \cite{Gersten_1975} to treat the Coulomb field effect. The main difficulty of CCSFA mentioned above is that the phase of the continuum wave function has a singularity near the core and the wave function cannot be straightforwardly applied in the calculation of the SFA matrix element. The singularity is removed using the matching procedure of the eikonal Coulomb-Volkov wave function with the bound atomic state. More recently, a new version of CCSFA has been derived more systematically in \cite{Torlina_2012,Torlina_2012b,Kaushal_2013}, rigorously implementing the matching procedure in the analytical R-matrix (ARM) theory.

The aim of this paper is twofold. Firstly, we extend CCSFA, considering high-order corrections to the eikonal wave function for the continuum electron, and employ it in CCSFA. We calculate PMD with the high-order CCSFA and discuss the impact of the corrections on PMD.  Secondly, we propose a method to avoid the Coulomb singularity in CCSFA amplitude without using the complex matching procedure of the ARM-theory. This is achieved calculating the SFA matrix element via the saddle-point integration method not only by time, but also by coordinate. When neglecting the high-order corrections, our method provides results which coincide with the ARM-theory.

The high-order Coulomb corrections to the eikonal Coulomb-Volkov wave function contain classical and quantum terms.
Why quantum correction to the eikonal Coulomb-Volkov wave function can be important? Recently a lot of experimental effort has been directed towards measuring the tunneling delay time  during the laser induced tunneling ionization \cite{Eckle_2008b,Pfeiffer_2012,Landsman_2014o}. The theoretical description of the tunneling delay time within a fully quantum theory is still missing. In the first order eikonal CCSFA the tunneling time is vanishing \cite{Torlina_2015}, because  the tunneling is described within Wentzel-Kramers-Brillouin (WKB) approximation, when the wave function under the barrier is real. Then usually a combined quantum-classical consideration is applied to describe  the tunneling delay time. The ionization is described quantum mechanically and the electron further propagation in continuum classically, see e.g. \cite{Pfeiffer_2012}. In the quasistatic regime of tunneling ionization the known Wigner formalism \cite{Wigner_1955} can be applied to calculate the tunneling  delay time as a time-derivative of the phase of the electron wave function under-the-barrier \cite{Yakaboylu_2013,Yakaboylu_2014b}. In the second step, the derived tunneling delay time is included into the initial conditions of the further classical propagation. However, in the nonadiabatic regime, when tunneling delay time is comparable with the laser period, the Wigner formalism is not applicable conceptually. In this case there is desire for systematic description of the modification of PMD due to the ionization delay time. In this context, the quantum corrections in our CCSFA  address the issue of the impact of Coulomb field of the atomic core on quantum effects during ionization.

Note that the quantum recoil effects for the continuum electron at photon emission and absorption in a strong laser field has been first considered in \cite{Avetissian_1997} (the relativistic version in \cite{Avetissian_1999}).  CCSFA based on this wave function was proposed in \cite{Avetissian_2001}, however, obtaining the final results  only in the Born approximation.

The structure of the paper is the following. In Sec.~\ref{Sec:parameters}   the considered system is introduced. The scheme of CCSFA is discussed in Sec.~\ref{Sec:SFA}. The results in the zeroth- and first-order SFA are presented in Secs.~\ref{Sec:Results} and \ref{Sec:Results1}. Comparison with the ARM- and Perelomov-Popov-Terent'ev (PPT)-theories are given in Secs.~\ref{Sec:ARM} and \ref{Sec:CCB}. Our main result - the second order SFA containing quantum corrections - is presented in Sec.~\ref{Sec:Results2}. The relation of the high-order CCSFA to the ionization delay time is analyzed in Sec.~\ref{Sec:Delays}, and to the heuristic quasiclassical theory of \cite{Klaiber_2015} in Sec.~\ref{Sec:exact}.

\section{The considered system}\label{Sec:parameters}

We consider the ionization process of an atom (ion) in a strong laser field. Our description is one-dimensional (1D). The active electron in the free atomic system is bounded by an 1D Coulomb potential
\begin{eqnarray}
V(x)=-\frac{Z}{|x|},
\end{eqnarray}
with the nuclear charge $Z$. Atomic units are used throughout. Initially the electron is in the ground state, which has the following asymptotic coordinate representation at $|x|\gg 1/\kappa$:
\begin{eqnarray}
\langle x|\phi(t)\rangle&=&\frac{\kappa
   (2\kappa  x)^{Z/\kappa }}{\sqrt{2Z \Gamma \left(\frac{2 Z}{\kappa }\right)}}\exp(-\kappa \left| x\right|+iI_pt)\\
	&\equiv& c_a\exp[S_{a}(x,t)],\nonumber\\
	S_{a}(x,t)&=&-\kappa \left| x\right|+iI_pt+Z/\kappa\log(2\kappa x),\\
	c_a&=&\frac{\kappa}{\sqrt{2Z \Gamma \left(\frac{2 Z}{\kappa }\right)}}
\end{eqnarray}
with $\kappa=\sqrt{2I_p}$ and $\log$ the logarithmic to basis $e$. Note that the ground state in the 1D Coulomb potential is antisymmetric.

In this paper we do not consider Coulomb effects at recollisions. Therefore, the ionization of the atom is considered in an half-cycle laser pulse, where the recollisions are excluded explicitly:
\begin{eqnarray}
F(t)=
\begin{cases}
E_0\cos(\omega t),\;\;{\rm for}\; |\omega t|< \pi/2\\
0,\;\;\;\; \;\;\;\;\; \;\;\;\;\;\;{\rm for}\; |\omega t|\geq \pi/2
\end{cases}
\end{eqnarray}
with the laser field amplitude $E_0$ and frequency $\omega$.

We consider the nonrelativistic regime of the interaction when the typical electron momenta in the bound state  as well as in the laser field are small  with respect to the speed of light $c$: $\kappa/c\ll 1$ and $E_0 /(c\omega)\ll 1$. We exclude also over-the-barrier ionization, which implies that the typical laser electric field $E_s$ is much smaller  with respect to the atomic field strength: $E_s/E_a<\kappa/16Z$, with $E_s=E_0$ in the tunneling ionization regime, $E_s=\gamma E_0$ in the multiphoton-ionization regime, $\gamma=\omega \kappa/E_0$ is the Keldysh parameter, and $E_a=\kappa^3$ is the atomic field. The depletion of the bound state is neglected. Finally, we assume that the photon energy is much less than the typical energies of the electron in the bound state $I_p$ and in the laser field $U_p$: $\omega\ll I_p,\,U_p$ with the ponderomotive potential $U_p=E_0^2/(4\omega^2)$, which are necessary for application of the saddle-point integration (SPI) method in calculation of the matrix element.
 Within these restrictions  we describe the ionization dynamics  analytically with SFA which will be explained in the following section.

\section{High-order Coulomb corrected strong-field approximation}\label{Sec:SFA}

The dynamics of the electron is described by  the Schr\"odinger-equation in the length gauge:
\begin{eqnarray}
i\partial_t|\psi(t)\rangle=H_0|\psi(t)\rangle-xF(t)|\psi(t)\rangle
\label{se}
\end{eqnarray}
with the unperturbed atomic Hamiltonian:
\begin{eqnarray}
H_0=\frac{\hat{p}_x^2}{2}+V(x).
\end{eqnarray}
We calculate PMD $w(p)=|M(p)|^2$ analytically via SFA-amplitude \cite{Becker_2002}:
\begin{eqnarray}
 M(p)=-i\int_{-\infty}^{\infty} dt\langle\psi_p(t)|xF(t)|\phi(t)\rangle,
\label{mp}
\end{eqnarray}
where $|\psi_p(t)\rangle$ is the solution of Eq.~(\ref{se}) with the asymptotic momentum $p$. The approximate solution in the high-order eikonal approximation is found using the following ansatz
\begin{eqnarray}
\langle x|\psi(t)\rangle\equiv\psi(x,t)=\frac{\exp[iS(x,t)]}{\sqrt{2\pi}}.
\end{eqnarray}
The latter is inserted into the Schr\"odinger equation which yields an equation for the eikonal function $S$:
\begin{eqnarray}
-\partial_t S=\frac{\left(\partial_x S\right)^2}{2}-xF+\alpha\left(V(x)-i\frac{\partial_{xx}{S}}{2}\right),
\label{qhj}
\end{eqnarray}
where we introduce an artificial perturbation parameter $\alpha$, which we will set unity later, such that we consider the Coulomb potential as well as the quantum correction perturbatively. Forseeing the subsequent calculation we can insert the typical value for the coordinate $x\sim\sqrt{\kappa/E_s}$ and time $t\sim \kappa/E_s$ into the perturbation 
of the original differential equation: $V\sim Z/x\sim Z\sqrt{E_s/\kappa}$ and $\partial_{xx}S\sim V t/x^2\sim Z\sqrt{E_s/\kappa}$
and see
that the quantum term is of the same order as the potential one and the simultaneous perturbative treatment of both of terms is justified when $E_0\ll E_a$, see equation (22) in \cite{Klaiber_2013a}.

In the usual eikonal approximation, in particular in \cite{Popruzhenko_2008a,Torlina_2012}, the last quantum term $\partial_{xx}{S}$ is neglected and the atomic potential is treated perturbatively in the eikonal equation (\ref{qhj}). In contrast to that we include into the consideration the quantum term, taking into account it, as well as the  atomic potential by perturbation theory. The quantum term yields a correction to the eikonal $S(x,t)$ of the second order. Therefore, we have to include in the solution of the eikonal also the second order correction due to the atomic potential.
With the ansatz
\begin{eqnarray}
S=S_0+\alpha S_1+\alpha^2S_2,
\end{eqnarray}
the zeroth, first and second order equations read:
\begin{eqnarray}
-\partial_t S_0&=&\frac{\left(\partial_x S_0\right)^2}{2}-xF\label{S0}\\
-\partial_t S_1&=&\partial_x S_0\partial_xS_1+V-i\frac{\partial_{xx}{S_0}}{2}\label{S1}\\
-\partial_t S_2&=&\frac{\left(\partial_x S_1\right)^2}{2}+\partial_x S_0\partial_xS_2-i\frac{\partial_{xx}{S_1}}{2}.
\label{S2}
\end{eqnarray}
The zeroth order equation is the Hamilton-Jacobi equation for the electron in the laser fields, which provides the well-known Volkov-action \cite{Volkov_1935},
\begin{eqnarray}
S_0(x,t)=\left[p+A(t)\right]x+\int^{t_f}_tdt'\frac{\left[p+A(t')\right]^2}{2}.
\end{eqnarray}
The first and second order equations are solved with the method of characteristics:
\begin{eqnarray}
S_1(x,t)&=&\int^{t_f}_tdt'V(x(t'))\nonumber\\
S_2(x,t)&=&\int^{t_f}_{t}dt'\frac{\left[\int^{t_f}_{t'}dt''\partial_xV(x(t''))\right]^2}{2}\nonumber \\
&-&i\int^{t_f}_{t}dt'\int^{t_f}_{t'}dt''\frac{\partial_{xx}V(x(t''))}{2},
\end{eqnarray}
where $x(t')=x+\int^{t'}_{t}ds[p+A(s)]$ is the electron classical trajectory in the laser field solely, and $A(t)$ is the laser vector potential,  with $F(t)=\partial_t A(t)$.

Further, we note that the terms in $S_0$, $S_1$ and the first term in $S_2$ are quasiclassical terms of order $1/\hbar$, the second summand in $S_2$ is a quantum term of order $\hbar^0$. The latter is a special feature of the 1D system. In a 3D-Coulomb system this term does not exist due to the fact that $\Delta V(r)=0$ for $r>0$.

We insert our approximate wave function for the continuum electron into the SFA amplitude of Eq.~(\ref{mp}). The two-dimensional integration in the matrix element is carried out by SPI method. For the latter we exponentiate the whole expression:
\begin{eqnarray}
M(p)&=&-\frac{ic_a}{\sqrt{2\pi}}\int dt dx \exp\{-iS^*(x,t)+\log[xF(t)]+S_a(x,t)\},\nonumber \\
\end{eqnarray}
where $^*$ indicates complex conjugation. The saddle-point conditions
\begin{eqnarray}
\left.\frac{d\zeta(x,t)}{dt}\right|_{(t,x)=(t_s,x_s)}&=&0\nonumber\\
\left.\frac{d\zeta(x,t)}{dx}\right|_{(t,x)=(t_s,x_s)}&=&0,
\end{eqnarray}
with $\zeta(x,t)=-iS^*(x,t)+\log[xF(t)]+S_a(x,t)$,  define the saddle-points $(t_s,x_s)$ around which the exponent is quadratically expanded in SPI. To be consistent with the expansion of $S(x,t)$, we  also expand the saddle-points $(t_s,x_s)$ up to second order:
\begin{eqnarray}
t_s&=&t^{(0)}_s+\alpha t^{(1)}_s+\alpha^2 t^{(2)}_s\nonumber\\
x_s&=&x^{(0)}_s+\alpha x^{(1)}_s+\alpha^2 x^{(2)}_s
\end{eqnarray}
and solve the saddle-point equations perturbatively. The corresponding zeroth, first and second order function in the exponent are
\begin{eqnarray}
\zeta_0(x,t)&=&-iS_0(x,t)+\log[x F(t)]+S_{a,0}(x,t)\nonumber\\
\zeta_1(x,t)&=&-iS_1(x,t)+S_{a,1}(x)\nonumber\\
\zeta_2(x,t)&=&-iS^*_2(x,t)
\end{eqnarray}
with $S_{a,0}(x,t)=-\kappa x+i I_pt$, $S_{a,1}(x)=Z/\kappa\log[2\kappa x]$.
Therefore, the zeroth order saddle-point equations read
\begin{eqnarray}
 -\partial_tS_0(x,t)&=&-I_p+i\frac{F'(t)}{F(t)} \\
 \partial_xS_0(x,t)&=&i\left(\kappa-\frac{1}{x}\right).
\end{eqnarray}
The zeroth order $(t^{(0)}_s,x^{(0)}_s)$ solution is found numerically. The higher order equations are  solved analytically and the solutions as well as the SFA amplitude $M(p)$ are expressed by $(t^{(0)}_s,x^{(0)}_s)$.

We may estimate  the zeroth order solution in a cosine-electric field as:
\begin{eqnarray}
t^{(0)}_s&\sim&\arcsin[i\kappa\omega/E_0]/\omega-i/\sqrt{\kappa E_s}\\
x^{(0)}_s &\sim&\sqrt{\kappa/E_s},
\end{eqnarray}
which correspond to the solutions in the case of a short-range potential. Note that the starting point of the ionization represented by the saddle-point in $x_s$ is far away from the atomic core, $x_s\kappa\gg 1$, therefore, the eikonal $S_1(x_s,t)$ is not singular.

The derivation of the higher order corrections to the saddle-points is straightforward, but cumbersome, and yields large analytical expression. We give only the structure of the first order solution of the $(t,x)$-saddle-points:
\begin{eqnarray}
t_s^{(1)}&=&\frac{-\partial_{xt}\zeta_0\partial_{x}\zeta_1+\partial_t\zeta_1\partial_{xx}\zeta_0}{\partial_{xt}\zeta_0^2-\partial_{tt}\zeta_0\partial_{xx}\zeta_0}\Big|_{x=x^{(0)}_s,t=t^{(0)}_s}\nonumber\\
x_s^{(1)}&=&\frac{\partial_{tt}\zeta_0\partial_{x}\zeta_1-\partial_t\zeta_1\partial_{xt}\zeta_0}{\partial_{xt}\zeta_0^2-\partial_{tt}\zeta_0\partial_{xx}\zeta_0}\Big|_{x=x^{(0)}_s,t=t^{(0)}_s}
\end{eqnarray}
The structure of the SFA amplitude up to second order is the following:
\begin{eqnarray}
&&M(p)\approx-i c_a \frac{\sqrt{2\pi}}{\sqrt{\det\zeta}}\\
&& \exp\left[\left(\zeta_0+\alpha\zeta_1+\alpha^2\zeta_2 \right.\right.\nonumber\\
&+&\left.\left.\left.\alpha^2 \frac{\partial_{xx}\zeta _0 \partial_t\zeta _1^2-2 \partial_x\zeta _1 \partial_{xt}\zeta _0 \partial_t\zeta _1 +\partial_{tt}\zeta _0 \partial_{x}\zeta _1^2}{2 \left(\partial_{xt}\zeta_0^2-\partial_{tt}\zeta _0 \partial_{xx}\zeta _0\right)}\right)\right|_{(x^{(0)}_s,t^{(0)}_s)}\right]\nonumber
\end{eqnarray}
where $\det\zeta$ is the Van-Vleck-Pauli-Morette~\cite{vanVleck} determinant of the matrix formed by the second order derivatives of $\zeta$:
\begin{eqnarray}
\det\zeta=\det\begin{pmatrix}
\partial_{xx}\zeta &\partial_{xt}\zeta\\
\partial_{tx}\zeta &\partial_{tt}\zeta \\
\end{pmatrix},
\end{eqnarray}
 $2\pi/\sqrt{\det\zeta}$ arises from SPI and represents intuitively  the typical size of the volume element $dxdt$.

Finally, we determine the maximum of PMD via the extremum condition
\begin{eqnarray}
\left.\frac{\partial M(p)}{\partial p}\right|_{p=p_m}=0,
\label{extremum}
\end{eqnarray}
which is solved again perturbatively $p_m=p^{(0)}_m+\alpha p^{(1)}_m+\alpha^2 p^{(2)}_m$, providing the maximum of the probability amplitude
\begin{eqnarray}
\left.M(p_m)\sim \frac{\exp(\zeta)}{\sqrt{\det\zeta}}\right|_{p=p_m},
\end{eqnarray}
with the function in the exponent being expanded up to second order. In the next section we will discuss the results of the calculations. The results obtained in the $n^{th}$-order expansion are referred as $S_n$-CCSFA.

\section{$S_0$-CCSFA}\label{Sec:Results}

The ionization amplitude in the zeroth order
\begin{eqnarray}
M(p)=\frac{\exp(\zeta_0)}{\sqrt{\det\zeta_0}},
\end{eqnarray}
corresponds to the standard SFA describing the ionization from a short-range potentials with $Z\ll \kappa$. As a check of accuracy for our SPI, we calculate analytically the $S_0$-CCSFA amplitude for a cosine-laser pulse and compare it with the PPT-result \cite{PPT}. The saddle-point for the most probable final momentum, i.e., the position and time where and when the ionization dynamics starts, can be given approximately analytically, where higher order terms in $E_0/E_a$ are dropped:
\begin{eqnarray}
x^{(0)}_s&\approx&\sqrt{\frac{\kappa}{E_s}}\nonumber\\
t^{(0)}_s&\approx&\frac{\arcsin[i\gamma]}{\omega}-\frac{i}{\sqrt{\kappa E_s}}.
\end{eqnarray}
The latter provides PMD  for ionization from a short-range potential in the leading terms in $E_0/E_a$:
\begin{eqnarray}
M(p)&=&\frac{\pi\kappa^2}{e E_s}\exp\left[-\frac{\kappa ^3 \left(-\sqrt{\gamma ^2+1} \gamma +2 \gamma ^2 \sinh ^{-1}\gamma +\sinh ^{-1}\gamma \right)}{2 \gamma ^3 E_0}\right. \nonumber\\
&-& \left. \frac{\left(p-p^{(0)}\right)^2}{\Delta ^2} \right],
\label{SFA0}
\end{eqnarray}
where $p^{(0)}=\int^{\infty}_0dtF(t)$ is the most probable momentum, $E_s=E_0\sqrt{1+\gamma^2}$ and
\begin{eqnarray}
\Delta =\frac{  \sqrt{E_s}}{\sqrt{  \kappa  \left(\sqrt{1+1/\gamma ^2} \sinh ^{-1}(\gamma )-1 \right)}}
\end{eqnarray}
is the width of the momentum distribution.
We note that the derived ionization amplitude differs from the PPT result in a short range potential by a constant  factor of $\pi/e\approx 1.16$, which arises due to the approximate $x$-integration with SPI in contrast to the exact $x$-integration in PPT.
The SPI error  mainly arises due to the Gaussian $x$-integration region $(-\infty,\infty)$ and can be reduced to a factor of $[1+{\rm erf}(1)]^2\pi/4e\approx0.98$, when  the integration spans only over the relevant region of coordinate $(0,\infty)$. In fact,  the region behind the atomic core ($x<0$) does not contribute to the ionization. In the high-order calculations the integration region will be restricted in the same way.

Generally, the saddle point approximation by time can be improved by including the third order term $\partial_{ttt}\zeta_0(x^{(0)}_s,t^{(0)}_s)(t-t^{(0)}_s)^3/6$ in the integration around the saddle point. However, the analysis of this term shows that it has no influence on the momentum distribution of the ionized electron, and changes the ionization probability only insignificantly due its relative smallness that can be estimated by $(E_0/E_a)/72$.

\section{$S_1$-CCSFA}\label{Sec:Results1}

The Coulomb field effect on PMD is described by the first order correction terms to the eikonal wave function. The correction that leads to a qualitative change compared to the short-range potential case is the first order Coulomb-correction $\zeta_1$ in the exponent:
\begin{eqnarray}
\frac{\exp(\zeta_0+\zeta_1)}{\sqrt{\det\zeta_0}}.
\label{zeta1}
\end{eqnarray}
Note that the preexponential term $\det\zeta_1$  yields a contribution which is small compared to the leading term in the order of $E_0/E_a$ and is neglected in  $S_1$-CCSFA.
This term is included in the wave function of the next order, and its effect will be discussed in $S_2$-CCSFA.

The Coulomb correction term in $S_1$-CCSFA, $\exp(\zeta_1)$, has two consequences. Firstly, $\exp(\zeta_1)$ changes the magnitude of the ionization probability via the following correction factor:
\begin{eqnarray}
&&\left|\frac{c_a}{c_{a,0}}\exp[\zeta_1(x^{(0)}_s,t^{(0)}_s)]\right|^2\approx \frac{4^{Z/\kappa } \left(\frac{1}{\sqrt[4]{\gamma ^2+1} \sqrt{f}}\right)^{\frac{2 Z}{\kappa }} }{\Gamma \left(\frac{2 Z}{\kappa }+1\right)}\\
&\times &\exp \left\{\frac{4 Z}{\kappa} \coth ^{-1}\left[\frac{\left(\sqrt{\gamma ^2+1}-1\right)}{\gamma } \coth \left(\frac{\sinh ^{-1}(\gamma )}{2} -\frac{\gamma  \sqrt{f}}{2\sqrt[4]{\gamma^2+1}}\right)\right]\right\}\nonumber
\end{eqnarray}
which yields  in the leading order in $E_0/E_a$:
\begin{eqnarray}
\left|\frac{c_a}{c_{a,0}}\exp[\zeta_1(x^{(0)}_s,t^{(0)}_s)]\right|^2\approx\frac{16^{Z/\kappa } f^{-\frac{2 Z}{\kappa }}}{\Gamma \left(\frac{2 Z}{\kappa }+1\right)}
\label{S1-factor}
\end{eqnarray}
with $f=E_0/\kappa^3$ and $c_{a,0}=\sqrt{\kappa/2\pi}$.  We note that the ionization amplitude of $S_0$-CCSFA Eq.~(\ref{SFA0}), with the correction factor of $S_1$-CCSFA Eq.~(\ref{S1-factor}), reproduces the PPT-ionization rate \cite{PPT}.

\begin{figure}
    \begin{center}
 \includegraphics[width=0.45\textwidth]{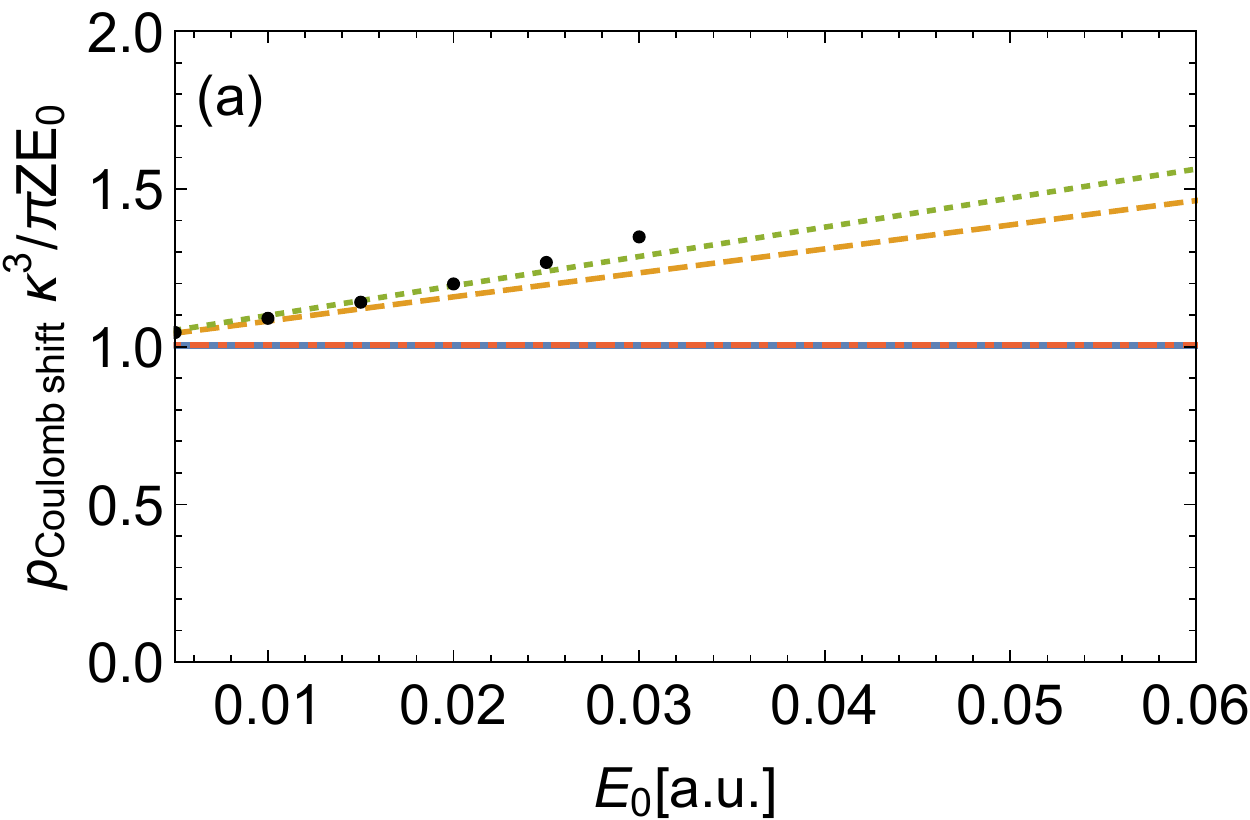}
  \includegraphics[width=0.45\textwidth]{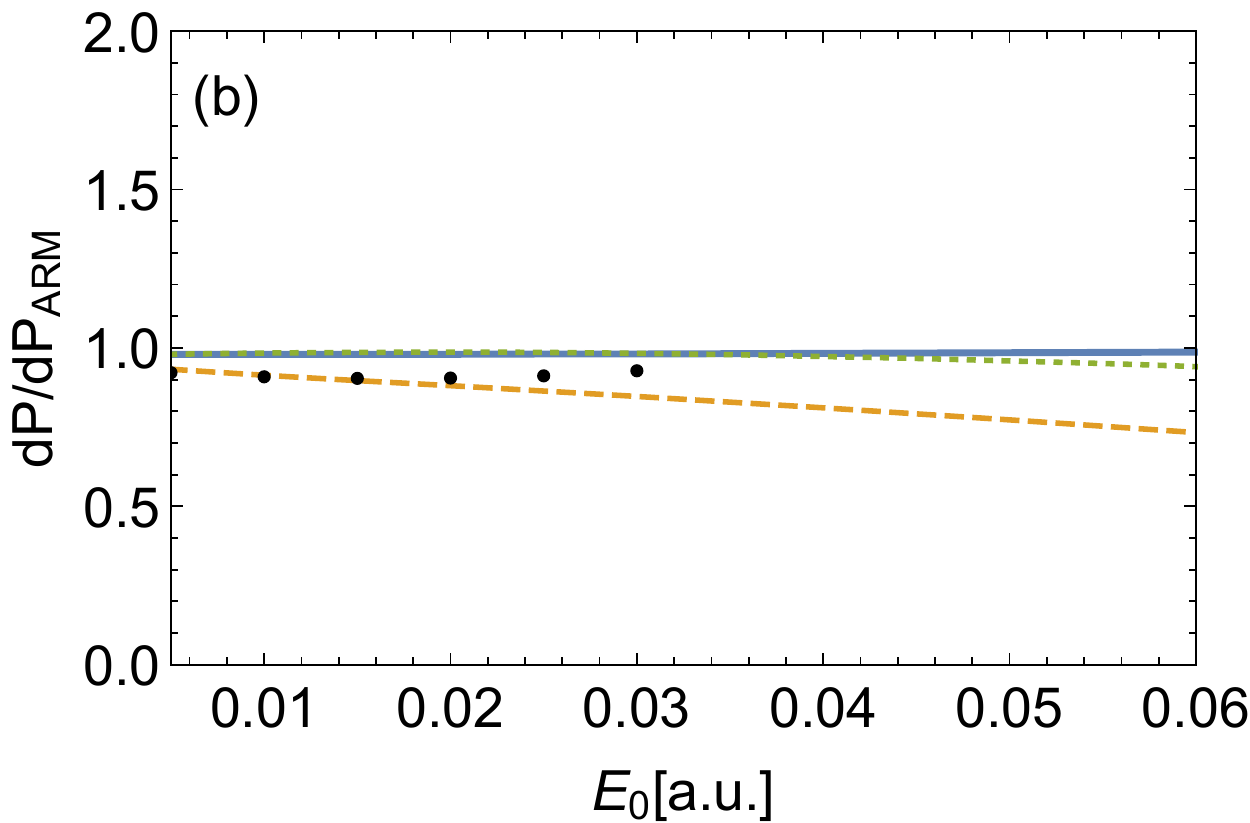}
           \caption{(color online) (a) The Coulomb-momentum shift vs the laser field strength in the quasistatic regime $\gamma=0.1$, $Z/\kappa=1$ with: (solid) the quasiclassical $S_1$-CCSFA, (dashed) the quasiclassical $S_2$-CCSFA, (dotted) the quantum $S_2$-CCSFA,  and (dotted-dashed) the ARM-theory. The black dots display the result of the method of Sec.~\ref{Sec:exact}.
	(b) The ratio of the ionization rate at the peak of the momentum distribution to the corresponding ARM-ionization rate in the quasistatic regime $\gamma=0.1$, $Z/\kappa=1$:
	(solid) for the quasiclassical $S_1$-CCSFA, (dashed) the quasiclassical $S_2$-CCSFA,  and (dotted)  the quantum $S_2$-CCSFA.  The black dots display the result of the method of Sec.~\ref{Sec:exact}.}
        \label{static}
    \end{center}
  \end{figure}
  \begin{figure}
    \begin{center}
 \includegraphics[width=0.45\textwidth]{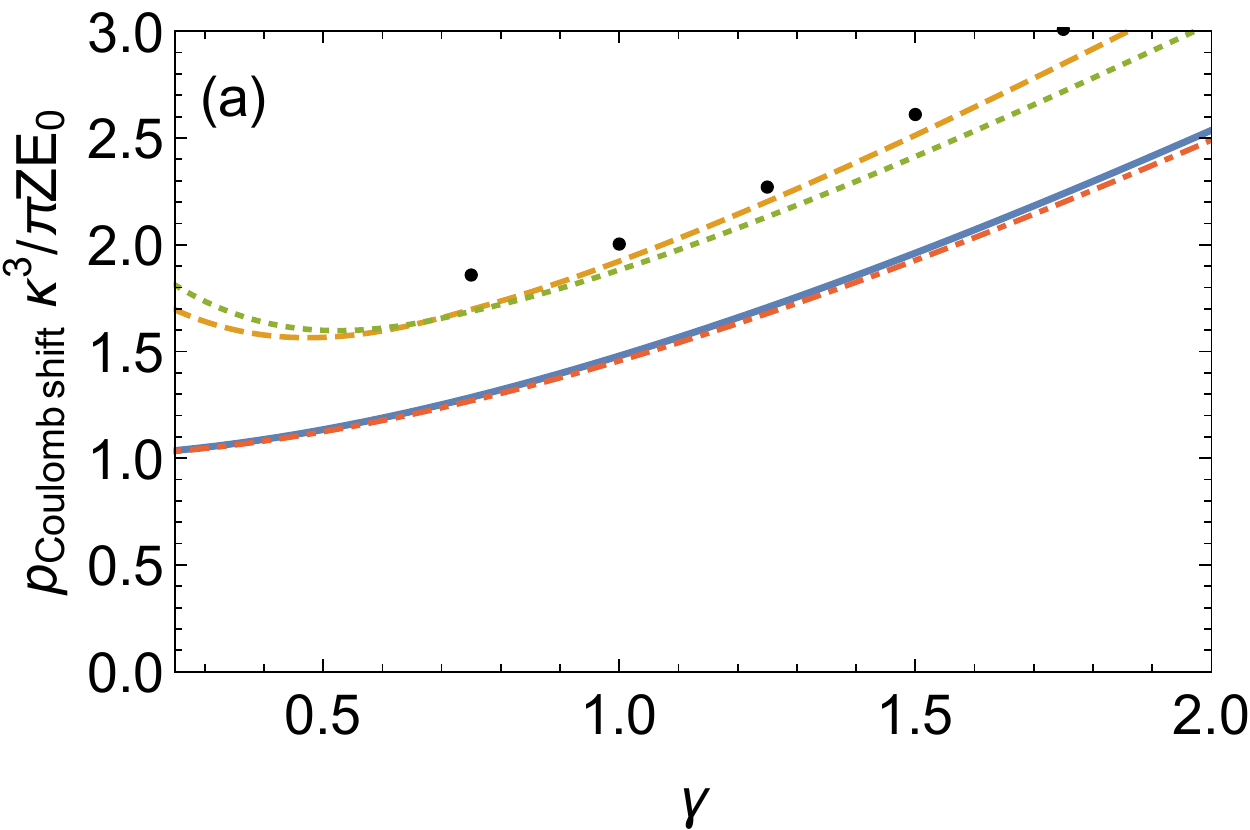}
  \includegraphics[width=0.45\textwidth]{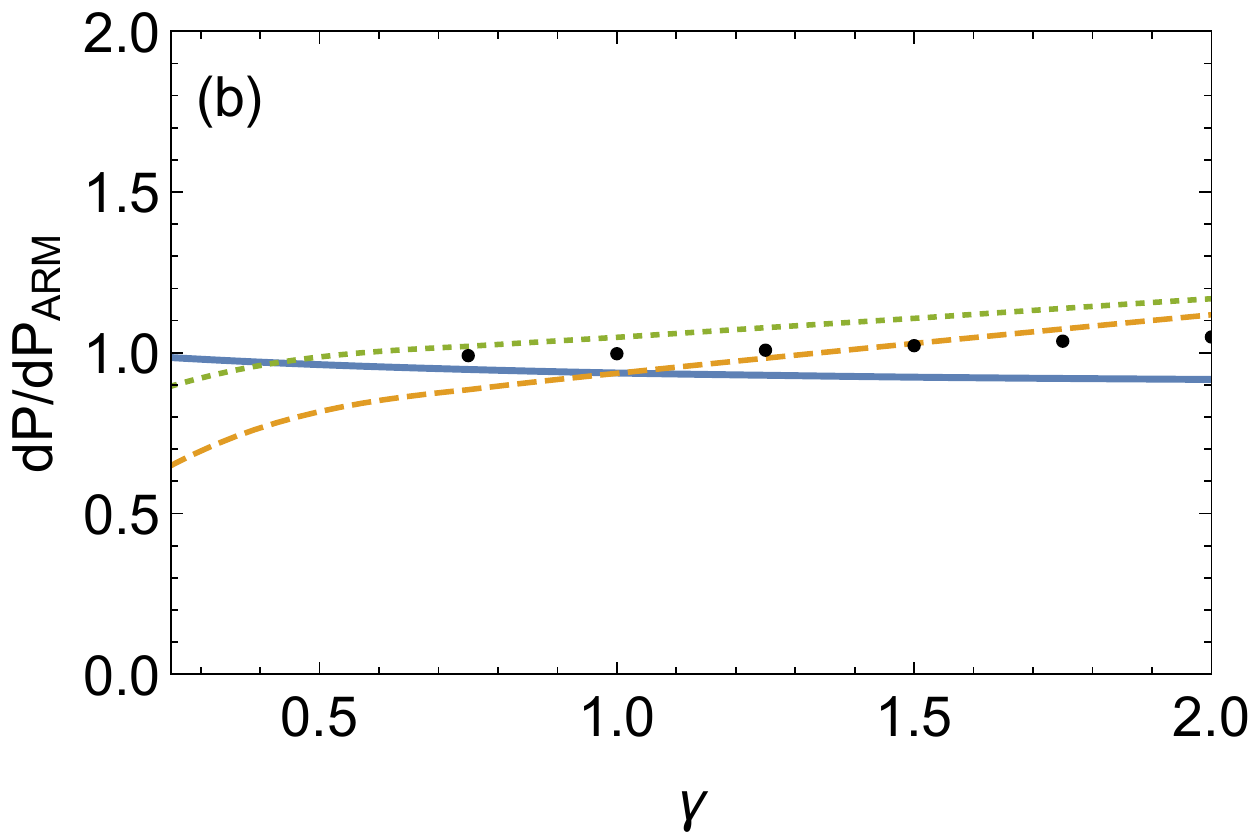}
           \caption{(color online) (a) The Coulomb-momentum shift vs the Keldysh-parameter $\gamma$ in the nonadiabatic regime $\omega=0.02$, $Z/\kappa=1$ with: (solid) the quasiclassical $S_1$-CCSFA, (dashed) the quasiclassical $S_2$-CCSFA, (dotted) the quantum $S_2$-CCSFA,  and (dotted-dashed) for the ARM-theory.
	(b) The ratio of the ionization rate at the peak of the momentum distribution to the corresponding ARM-ionization rate vs the Keldysh-parameter $\gamma$ in the nonadiabatic regime $\omega=0.02$, $Z/\kappa=1$: (solid) for the quasiclassical $S_1$-CCSFA,  (dashed) the quasiclassical $S_2$-CCSFA,  and  (dotted) the quantum $S_2$-CCSFA. The black dots display the result of the method of Sec.~\ref{Sec:exact}.}
        \label{na}
    \end{center}
  \end{figure}

Secondly, $\zeta_1$ yields a shift of the momentum distribution due to a momentum transfer to the Coulomb potential during the motion of the ionized electron in the continuum immediately after leaving the tunnel exit (we underline  again that here recollisions are not considered). The momentum shift derived from the condition of the extremum of  $M(p)$, Eq.~(\ref{extremum}), with $S_1$-CCSFA, is shown in Fig.~\ref{static}. We can give also an analytical estimation of the momentum shift via $\partial_x\zeta_1$, which in the static regime, $\gamma\ll 1$, is:
\begin{eqnarray}
\Delta p_C\approx{\rm Re}\left\{{\partial_x\zeta_1(x^{(0)}_s,t^{((0)}_s)}\right\}\approx\frac{\pi ZE_0}{\kappa^3},
\label{CMstatic}
\end{eqnarray}
and in the nonadiabatic regime, $\gamma\gtrsim 1$, it is:
\begin{eqnarray}
\Delta p_C\approx{\rm Re}\left\{{\partial_x\zeta_1(x^{(0)}_s,t^{((0)}_s)}\right\}\approx\frac{\gamma^2 ZE_0}{\kappa^3},
\label{CMna}
\end{eqnarray}
where in the latter the electron trajectory $x(t)\approx x_e+E_0t/\omega$ is used taking into account that the drift during the half-cycle pulse from $t=0$ up to $T=\pi/\omega$ is small compared to the tunnel exit $x_e$ and can be dropped. The coordinate of the tunnel exit [$x_e=2/\gamma^2(\sqrt{1+\gamma^2}-1)I_p/E_0$] in the nonadiabatic case, $\gamma\gg 1$, is $x_e\approx \kappa/\omega$, see \cite{Klaiber_2015}.

In Figs.~\ref{static}(a) and  \ref{na}(a) one can observe that the Coulomb momentum shift values estimated by Eqs.~(\ref{CMstatic}) and (\ref{CMna}) almost exactly corresponds to the $S_1$-CCSFA theory.
Physically this  result can be interpreted as  a verification of the simple-man model \cite{Simpleman}, where instantaneous tunneling up to the exit $x_e$ is followed by classical propagation in the continuum, where the Coulomb field of the atomic core induces a momentum shift
\begin{eqnarray}
\Delta p_C\approx -\int_{0}^{\infty}dt\partial_x V(x(t)).
\label{CM}
\end{eqnarray}
The latter expression yields Eqs.~(\ref{CMstatic}) and (\ref{CMna}) when the electron trajectory $x(t)$  is used with either static or nonadiabatic tunnel exit coordinate.
In the nonadiabatic regime the Coulomb momentum shift is larger than in the quasistatic case, see Fig.~\ref{na}, because the nonadiabatic  trajectory is close to the atomic core for a longer time interval.

The approach of $S_1$-CCSFA is physically equivalent to the ARM-theory. The only difference is in how the Coulomb singularity is treated. While in the ARM-theory a rigorous matching of the electron wave function in the continuum to the bound state is employed, in the $S_1$-CCSFA the Coulomb singularity is avoided simply using additional SPI for the coordinate integration. In the next section we provide in details the comparison of  $S_1$-CCSFA with ARM.

\section{Comparison $S_1$-CCSFA with ARM-theory}\label{Sec:ARM}

We provide a comparison of the ARM-theory \cite{Torlina_2012} with $S_1$-CCSFA in Figs.~\ref{static} and \ref{na}, where the most probable momentum and the corresponding rate are shown.  The figures indicate that the results of $S_1$-CCSFA and ARM  for the most probable momentum, as well as for the rate are mostly identical. There is only slight difference in the most probable momentum and in the ionization rate in the nonadiabatic regime at large $\gamma$. To understand why this slight difference arises, let us look into details. In the derivation of the ionization amplitude in the ARM-theory  one arrives at the following expression for the amplitude:
\begin{eqnarray}
M^{ARM}(p)=-i \int dt \frac{\kappa c_a}{\sqrt{2\pi}}\exp[-i S_0(b,t)-i S_1(b,t)+S_a(b,t)]\nonumber \\
\end{eqnarray}
(this equation is the 1D analogue of Eq.~(28) of \cite{Torlina_2012}) where $b$ is the matching point of the bound  and the continuum states, and the amplitude is approximately independent of  the parameter $b$. The latter implies that the exponent in the expression fulfils the SPI condition at $b$.
Using SPI for the time-integration  yields
\begin{eqnarray}
M^{ARM}(p)&\approx &-i \kappa c_a\sqrt{\frac{1}{-\partial_{tt}S_0(b,t_s)}}\nonumber \\
&\times & \exp[-i S_0(b,t_s)-i S_1(b,t_s)+S_a(b,t_s)]\label{ARM2}
\end{eqnarray}
with $-\partial_{tt}S_0(b,t_s)\approx\kappa E_s$.

On the other side, the SPI over the time and coordinate in our $S_1$-CCSFA yields:
 \begin{eqnarray}
M(p)&\approx & -i  c_a x^{(0)}_s F(t^{(0)}_s)\frac{\sqrt{2\pi}}{\sqrt{-\left.\det\zeta\right|_{(x^{(0)}_s,t^{(0)}_s)}}} \label{ourS1}\\
&\times &\exp[-i S_0(x^{(0)}_s,t^{(0)}_s)-i S_1(x^{(0)}_s,t^{(0)}_s)+S_a(x^{(0)}_s,t^{(0)}_s)],\nonumber
\end{eqnarray}
where $\left.-\det\zeta\right|_{x^{(0)}_s,t^{(0)}_s}\approx  2E_s^2$.

For comparison of Eqs.~(\ref{ARM2}) and (\ref{ourS1}) we use $b=x^{(0)}_s$ as the ARM-amplitude does not depend on the matching point within the barrier near the condition of the coordinate SPI.

In the further derivation of the final ARM-expression in \cite{Torlina_2012} the factor $\exp(-E_s b^2/2\kappa)$ is neglected and after this operation the SFA amplitude, estimated for the typical values for $x^{(0)}_s\sim\sqrt{\kappa/E_s}$ and $F(t^{(0)}_s)\sim E_s$, differs from the ARM amplitude by a constant  factor $\sqrt{\pi/e}$, which is close to unity.

Thus, the reason for a small difference between the ARM- and $S_1$-CCSFA theories is that in the $S_1$-CCSFA  SPI with respect to the coordinate is applied, which implies that higher order derivatives with respect to $x$ are neglected. Meanwhile, in the ARM-theory a term $\exp(-E_s b^2/2\kappa)$ is neglected which is of the same order. Therefore, the ARM- and $S_1$-CCSFA theories are of same accuracy.

\section{On the PPT Coulomb correction factor in nonadiabatic regime}\label{Sec:CCB}

We stated in Sec.~\ref{Sec:Results1} that $S_1$-CCSFA provides Coulomb correction factor for the ionization amplitude which coincides with the PPT theory \cite{PPT}. Recently, a modification for the Coulomb correction factor is calculated in the nonadiabatic regime within PPT theory \cite{Popruzhenko_2008c}. The modification  includes the effect of frustrated ionization \cite{Nubbemeyer_2008,Eichmann_2009}, i.e., the capture of low energy electrons in the Coulomb potential of the atomic core after switching off the laser field.

To elucidate the role of the frustrated ionization, we derive this factor heuristically with the help of SFA. In the asymptotic PMD after switching off the laser field only electrons which gain sufficient energy in the laser field can leave the Coulomb potential. This energy gain depends on the ionization time, $ \varepsilon_e\sim A(t_i)^2/2$, with the ionization time $t_i$, and the laser vector potential of the sinusoidal field $A(t)=E_0/\omega \sin(\omega t)$, and has to be larger than the negative Coulomb  energy: $ \varepsilon_e>Z/x_e=Z/\kappa\omega$. In the latter we take into account that the minimum of the asymptotic coordinate cannot be smaller than the tunnel exit coordinate. From this it follows that electrons that tunnel close to the peak of the laser field are captured by the Coulomb potential and will not be detected. Only electrons with a certain ionization time away from the peak will be ionized. The reduction factor  of the ionization rate due to the capturing can therefore be estimated via Eq.~(\ref{SFA0}) at $\gamma\gg 1$:
\begin{eqnarray}
	 \left|\frac{M(-A(t_i))}{M(0)}\right|^2	\approx \left(\frac{2\gamma}{e}\right)^{-2 Z/\kappa},
\end{eqnarray}
which coincides  with the additional factor derived in \cite{Popruzhenko_2008c}. As our calculations does not include recollisions and the effect of the frustrated ionization, this factor is not relevant to our result.

\section{$S_2$-CCSFA}\label{Sec:Results2}

The $S_1$-CCSFA considered up to now provides results of PPT- and ARM-theories, circumventing the necessity of the wave functions matching procedure. The coincidence of the results is due to the fact that the saddle-point  of the  coordinate SPI is rather far from the atomic core, where the eikonal wave function for the electron is still valid. In this section we account for high-order corrections in CCSFA approach, to go beyond the known results of PPT- and ARM-theories.

The $S_2$-CCSFA contains as a quasiclassical correction term ($\sim \hbar^0$) as well as quantum correction terms ($\sim \hbar$). One quantum correction term is in the $S_2$-term of the eikonal, and  the second is
in the ionization amplitude due to the $S_1$-term in the prefactor (determinant), which has been neglected in $S_1$-CCSFA because of smallness, see Eq.~(\ref{zeta1}).
The SFA that includes only the quasiclassical correction term in the second order will be called quasiclassical $S_2$-CCSFA, whereas the CCSFA with all correction terms - the quantum $S_2$-CCSFA.

The second order corrections to the ionization amplitude are small and change the momentum distribution only quantitatively. The shift of the peak of PMD and the change of probability at the peak of the momentum distribution  due to these terms  are displayed for the quasistatic regime in Fig.~\ref{static}, and for the nonadiabatic regime in Fig.~\ref{na}. In both  regimes the second order correction terms do not change the ionization probability significantly, but increase the Coulomb momentum shift  compared to the $S_1$-CCSFA result.

The three different correction terms have a distinct physical origin. The second order terms in the quasiclassical  $S_2$-CCSFA decrease the ionization probability and increase the momentum shift. These changes are due to the decreases of the effective potential barrier formed by  the Coulomb field of the atomic core and the laser field. In fact, the tunnel exit coordinate taking into account the Coulomb field can be found from the relation:
\begin{eqnarray}
-I_p=-E_0x-\frac{Z}{|x|},
\end{eqnarray}
which can be solved exactly in $x$. An expansion of this solution in $E_0/E_a$ gives
\begin{eqnarray}
x_e\approx \frac{I_p}{E_0}\left(1-\frac{4ZE_0}{\kappa E_a} \right).
\end{eqnarray}
The $S_1$-CCSFA contains the simpleman exit $x_e\approx I_p/E_0$, while in $S_2$-CCSFA  the tunnel exit shifts closer to the atomic core due to the Coulomb correction according to the term of the order of $E_0/E_a$. This effect increases the Coulomb momentum shift in the continuum according to Eq.~(\ref{CM}).

Due to the Coulomb field effect according to the second order quasiclassical corrections, the tunneling probability decreases because of larger damping from the tunneling exponent $\exp(-\int |p(x)|dx)$, with $p(x)=i\sqrt{2(I_p-x E_0+Z/|x|)}$. The same can be deduced from the SFA formalism. The decrease of the ionization probability is mainly due to the next to leading order $E_0/E_a$ corrections in the Coulomb-correction factor neglected in the $S_1$-CCSFA (the second term in the bracket):
\begin{eqnarray}
\left|\frac{c_a}{c_{a,0}}\exp[\zeta_1(x^{(0)}_s,t^{(0)}_s)]\right|^2\approx
\frac{1}{2} \left(\frac{4 \kappa ^3}{E_0}-2 \kappa  x^{(0)}_s\right)^{\frac{2
   Z}{\kappa }}.
\end{eqnarray}
From this formula one can observe that the Coulomb correction factor decreases with an increasing saddle-point $x^{(0)}_s$, which fits to the intuitive explanation above.

The quantum correction to  $S_2$ increases the tunneling probability, that can be understood intuitively as decrease of the tunneling barrier. In fact, the quantum correction term $-i\partial_{xx}S/2\sim -ip'(x)$  in Eq.~(\ref{qhj}) is equivalent to an additional term in the effective potential $V_{eff}=V-x F(t)-i\partial_{xx}S/2$, which decreases the effective potential and consequently the coordinate of the tunnel exit. From this it follows that quantum corrections increase the ionization probability and the Coulomb momentum shift in the continuum motion.

\section{Relation to the ionization delay time}\label{Sec:Delays}

\begin{figure}
    \begin{center}
 \includegraphics[width=0.45\textwidth]{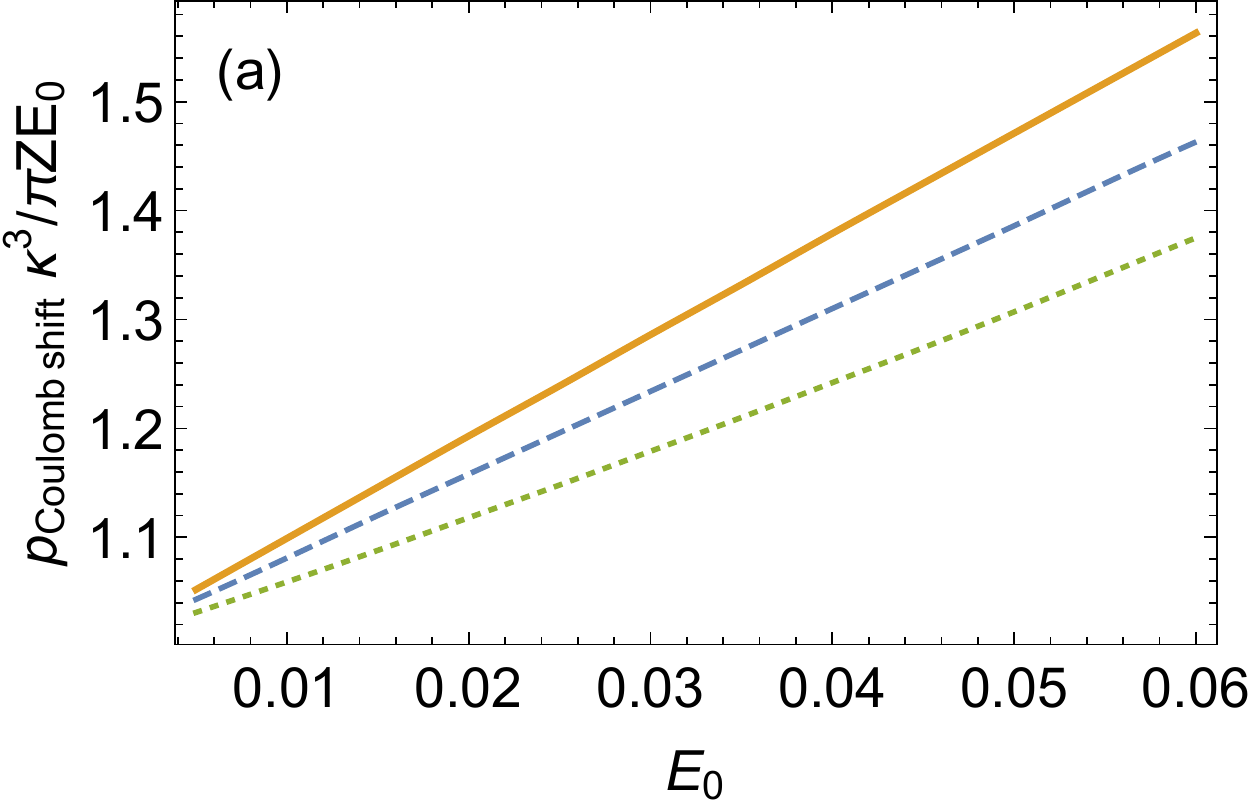}
  \includegraphics[width=0.45\textwidth]{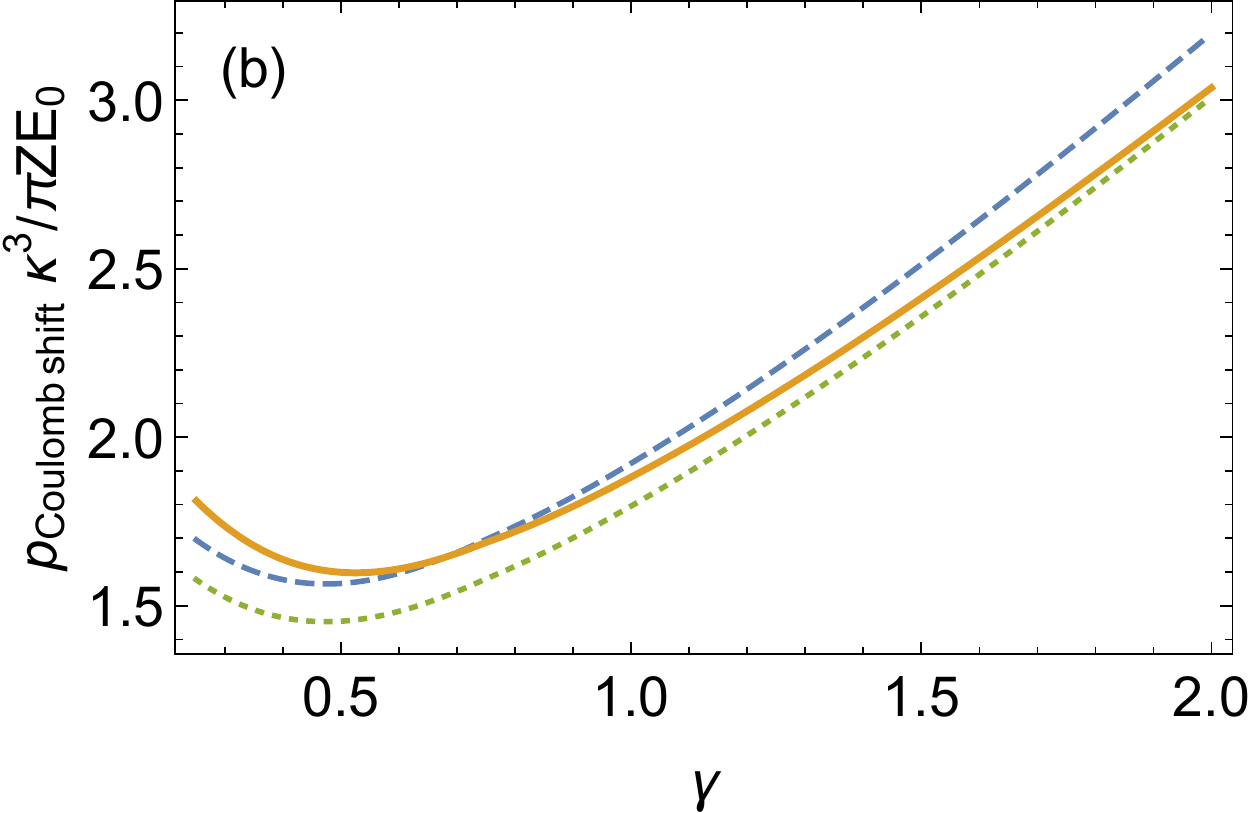}
           \caption{(color online) (a) The Coulomb momentum shift of the final momentum vs the laser  field in the quasistatic regime of $\gamma=0.1$: (dashed) via classical $S_2$-CCSFA, (solid) via quantum $S_2$-CCSFA including both quantum corrections, (dotted) via quantum SFA where the quantum corrections in the exponent are dropped, (b) The Coulomb momentum shift of the final momentum vs $\gamma$ in the nonadiabatic regime of $\omega=0.02$ a.u.: (dashed) via classical $S_2$-CCSFA, (solid) via quantum $S_2$-CCSFA including 	both quantum corrections, (dotted) via quantum SFA where quantum corrections in the exponent are dropped. }
        \label{tt}
    \end{center}
  \end{figure}

Let us  inspect the role of the different Coulomb quantum correction terms in the ionization amplitude: the quantum term in $S_2$, and the the correction term in the determinant due to high-order terms originating from $S_1$. We observe that for the Coulomb momentum shift in both regimes the quasiclassical $S_2$- and quantum $S_2$-curves are very close to each other, see  Figs.~\ref{static}(a) and \ref{na}(a). This indicates that  the two quantum corrections almost compensate each other. Whereas the quantum term in $S_2$ increases the momentum shift, the correction term in the determinant decreases it, yielding to an approximately net zero change. The role of the quantum correction due to the determinant term is further clarified in Fig.~\ref{tt}.  The compensation is different in the quasistatic and nonadiabatic regimes. While in the quasistatic regime the overall momentum shift is positive (the determinant term contribution is less important), in the nonadiabatic regime  the net momentum  corrections is negative (the determinant term contribution is more conspicuous).

Physically the momentum shifts can be interpreted as a delay time at the detector in the atoclock type setup \cite{Eckle_2008b} with respect to the simple-man model prediction. The quantum correction term in $S_2$ induces a positive delay time, and the determinant term negative delay time  of the same order in comparison to the simple-man result given by the quasiclassical $S_2$-CCSFA.
We underline that the delay time due to the Coulomb quantum corrections is an additional effect on top of the Wigner delay time \cite{Wigner_1955} at tunnleing ionization \cite{Yakaboylu_2014b}. The latter is not described by CCSFA.

In the more realistic 3D-case the quantum correction in $S_2$ is vanishing as the term $\Delta V(r)=0$ in Eq.~(\ref{S1}) for the 3D Coulomb potential. Then, in 3D case the overall delay time due to Coulomb quantum corrections will be connected only with the determinant term and, consequently,  negative.  Moreover, one can show that the time-derivatives of $S(x,t)$ in the determinant are responsible for the negative delay time, whereas spatial derivatives play a minor role for this effect. This indicates that the negative delay time due to Coulomb quantum corrections is not connected with the spatial uncertainty of the bound state, but an effect due to quantum corrections in the continuum state.
%This hints that the  delay time due to Coulomb quantum corrections can be related to the effect of the Coulomb field on the Wigner delay time \cite{Wigner_1955}.
Furthermore, one observes from Fig.~\ref{tt} that the delay time effect increases in the near threshold regime (large $E/E_a$), whereas in the deep-tunneling regime it is not significant. These are in line with the specific properties of the tunneling ionization delay time \cite{Yakaboylu_2014b}.

\section{Comparison with the heuristic quasiclassical method}\label{Sec:exact}

Finally, we discuss the relation of the results of our systematic $S_n$-CCSFA with the heuristic quasiclassical approach (HQA) of Ref. \cite{Klaiber_2015} for nonperturbative treatment  of Coulomb field effects during the under-the-barrier motion in strong field ionization.

Briefly recalling HQA, we begin with the ionization probability expressed via the quasiclassical propagator:
\begin{eqnarray}
M(p)& = & -i\int dx_f dx dt \exp(-ipx_f)G(x_f,x,t_f,t)\nonumber \\
&\times & xF(t)\frac{\phi_0(x,t)}{\sqrt{2\pi}},
\label{mp2}
\end{eqnarray}
with the quasiclassical Green's function
\begin{eqnarray}
G(x_f,x,t_f,t)=\sqrt{\frac{\partial_{x_f,x}\tilde{S}_c}{2\pi}}\exp(i\tilde{S}_c),
\end{eqnarray}
and the quasiclassical action, evaluated along the classical trajectory corresponding to the most probable electron,
\begin{eqnarray}
\tilde{S}_c=\int^{t_f}_{t} dt\left(\frac{\dot{x}(t)^2}{2}+xF(t)-V(x)\right).
\end{eqnarray}
The $x_f$ integral in Eq.~(\ref{mp2}) can be calculated via SPI, yielding
\begin{eqnarray}
M(p)=-\frac{i}{\sqrt{2\pi}} \int dx dt \exp(-i p x_f+i \tilde{S}_c)xF(t)\phi(x,t),\nonumber \\
\label{Mp}
\end{eqnarray}
where $\partial_{x_f,x}\tilde{S}_c/\partial_{x_f}^2\tilde{S}_c\approx 1$ was used.
The quasiclassical action fulfills the Hamilton-Jacobi equation:
\begin{eqnarray}
-\partial_t \tilde{S}_c=\frac{\left(\partial_x \tilde{S}_c\right)^2}{2}-x F(t)+V(x).
\end{eqnarray}
The saddle-point equations that occur when the $x,t$-integral in Eq.~(\ref{Mp}) is evaluated, are
\begin{eqnarray}
\partial_x \tilde{S}_c&=&i\kappa-\frac{(Z+\kappa)i}{\kappa x} \label{ps}\\
-\partial_t \tilde{S}_c&=&-\kappa^2/2+i\frac{F'(t)}{F(t)},
\end{eqnarray}
which also define the initial momentum and energy of the ionizing electron.
These two equations are inserted into the Hamilton-Jacobi equation yielding a new defining equation for the saddle-points:
\begin{eqnarray}
\frac{2 i \kappa  x_s F'(t_s)}{F(t_s)}+2 \kappa  x_s^2 F(t_s)+\frac{\kappa ^2-2 \kappa ^3 x_s+Z^2+2 \kappa  Z}{\kappa  x_s}=0.\nonumber \\
\end{eqnarray}
The latter can be simplified in the limit of $F(t_s)\ll\kappa^3$:
\begin{eqnarray}
x_s\approx \sqrt{\frac{\kappa}{F(t_s)}}.\label{xs}
\end{eqnarray}

The  under-the-barrier trajectory is derived solving Newton equations in the laser and  Coulomb fields numerically for different $t_s$, with the initial coordinate and momentum as a function of the saddle time $t_s$ according to Eqs.~(\ref{xs}) and (\ref{ps}). Since we are interested in the peak of the final momentum distribution, the most probable trajectory has to be found, corresponding to a specific $t_s$. This is accomplished via the additional conditions  defining the tunnel exit.

Firstly, for the most probable trajectory the coordinate should become real at the tunnel exit ${\rm Im} \{x(t_e)\}=0$, and secondly, the electron velocity along the tunneling direction should vanish  $\dot{x}(t_e)=0$. With these boundary conditions the Coulomb-corrected exit $x(t_e)$  is then  deduced from the solutions of the Newton equations and also the asymptotic final momentum $p=\dot{x}(t_f)$ is derived.

It is true that different integration contours can be chosen without changing the final electron momentum. For different integration contours the trajectory of the electron far from the exit is the same, while in  close vicinity of the exit there are still small deviations. Our choice of the contour is based on the physical condition that the electron  momentum along the tunneling direction should be vanishing at the tunnel exit. In this way the most probable trajectory is determined.

The results of HQA are displayed as dots in Fig.~\ref{static} and~\ref{na}. They are in accordance with the classical $S_2$-CCSFA results. For strong fields, i.e. larger $E_s/E_a$, HQA gives slightly larger momentum shifts and ionization probabilities. This is due to the fact that in the exact treatment of the Coulomb potential in HQA, the tunneling barrier is smaller than the barrier in the quasiclassical $S_2$-CCSFA.

Further we want to note that in the quasi-classical treatments presented here, i.e. in the quasi-classical $S_0$-, $S_1$-, $S_2$-SFAs as well as in the HQA the saddle points in time and in coordinate for the most proabable momentum fulfill ${\rm Re}\{t_s(p_m)\}=0$ and  ${\rm Im}\{x_s(p_m)\}=0$ in all considered regimes. This indicates that in a quasi-classical description tunneling happens instantaneous and that there  exists neither a static nor a nonadiabatic tunnel-ionization induced time delay with respect to the laser field maximum.

\section{Conclusion}

We have investigated the role of high-order corrections in the eikonal CCSFA. There are quasiclassical and quantum second order corrections to CCSFA. The second order terms in the quasiclassical  $S_2$-CCSFA decrease the ionization probability and increase the momentum shift. These changes are due to the decreases of the effective tunneling potential barrier of ionization when the Coulomb field of the atomic core is accounted for perturbatively.

There are two types of second order quantum correction terms which originate either from the $S_2$ in the eikonal, or the correction term in the prefactor due to high-order terms stemming from $S_1$. The first term is specific for 1D problem and absent in 3D theory.  The quantum term in $S_2$ increases the momentum shift, the correction term in the prefactor decreases it, yielding to a compensation. However, the compensation is different in the quasistatic and nonadiabatic regimes. While in the quasistatic regime the overall momentum shift is positive, in the nonadiabatic regime  the net momentum  corrections is negative.

Relating the momentum shift to the ionization delay time at the detector as in the attoclock  setup   and taking into account that in 3D case  the quantum correction in $S_2$ is not present, we may conclude that in 3D case the variation of the delay time due to Coulomb field effect will be  negative due to the solely determinant correction terms.  The fact that  time-derivatives in the determinant are responsible for the negative delay time points out that it  can be related to the Wigner delay time. Further, one observes that the delay time effect increases in the near threshold regime, whereas in the deep-tunneling regime it is not significant. This property is characteristic for the tunneling Wigner delay time.

Our approach for CCSFA in the first order approximation coincides with the ARM-theory and demonstrates a simple method to cope with the Coulomb singularity and circumvent the  matching procedure of the ARM-theory by means of the saddle-point integration of the  amplitude not only by time but also by coordinate.

The comparison of our heuristic quasiclassical approach  \cite{Klaiber_2015} for treating exactly the Coulomb effects for the electron under-the-barrier dynamics during tunneling ionization, with the quasiclassical second order CCSFA shows that the heuristic  approach  gives slightly larger momentum shifts and ionization probabilities. This stems from a more accurate description of the  tunneling barrier in the quasiclassical heuristic method.

\bibliography{strong_fields_bibliography}

\end{document}